\documentclass[prd,twocolumn,amssymb,12,showpacs]{revtex4-1}

\pdfoutput=1
\usepackage{units}
\usepackage{xcolor}
\usepackage{amsmath}
\usepackage{amssymb}
\usepackage{stmaryrd}
\usepackage{esint}
\usepackage{filecontents}
\usepackage{siunitx}
\usepackage{comment}
\usepackage[bottom]{footmisc}
\usepackage{diagbox}
\usepackage{float}
\usepackage[justification=raggedright,
  singlelinecheck=on]{caption}
\restylefloat{table}
\usepackage[unicode=true,pdfusetitle,
 bookmarks=true,bookmarksnumbered=false,bookmarksopen=false,
 breaklinks=false,pdfborder={0 0 0},backref=true,colorlinks=true,linkcolor=blue!50!black,urlcolor=red!50!black,citecolor=blue!50!black]
 {hyperref}
\usepackage{breakurl}
\begin{document}	
\title{Gravitational Waves and Degrees of Freedom in Higher Derivative Gravity}
\author{Patric H\"olscher} \email{patric.hoelscher@physik.uni-bielefeld.de} \affiliation{Fakult\"at f\"ur Physik, Bielefeld University, Postfach 100131, 33501 Bielefeld, Germany} 

\begin{abstract}
We study the degrees of freedom of the metric in a general class of higher derivative gravity models, which are interesting in the context of quantum gravity as they are (super)renormalizable. 
First, we linearize the theory for a flat background metric in Teyssandier gauge for an arbitrary number of spacetime dimensions $D$. The higher-order derivative field equations for the metric perturbation can be decomposed into tensorial and scalar field equations resembling massless and massive wave equations. For the massive tensor field in $D$-dimensions we demonstrate that the harmonic gauge condition is induced dynamically and only the transverse modes are excited in the presence of a matter source. For the special case of quadratic gravity in four-dimensional spacetime, we show that only the quadrupole moment contributes to the gravitational radiation from an idealized binary system.
\end{abstract}

\maketitle

\section{Introduction\label{sec:Introduction}}

General relativity (GR) as the standard theory of gravity works very well on solar system distance scales \cite{Will2014}. 
Nevertheless, considering only the luminous matter, GR cannot explain several astrophysical \cite{2017_DelPopolo_SmallScaleProblemsoflambdacdm} as well as cosmological \cite{2016_Bull_BeyondCDMProblemsSolutionsandtheRoadAhead_PotDU} phenomena. This led to the introduction of dark matter as a new particle, which interacts only very weakly with other standard model particles, but couples to gravity. Besides that, the small value of the cosmological constant (compared to the zero-point energy density from particle physics) is not understood \cite{1989_Weinberg_TheCosmologicalConstantProblem_Romp,2001_Carroll_TheCosmologicalConstant_LRR,2008_Bousso_TASILecturesontheCosmologicalConstant_GRG}.

Further, it seems to be impossible to combine GR with quantum mechanics in the ultraviolet regime. A perturbative quantization of GR leads to divergences, which cannot be renormalized \cite{1974_Veltman_GtHooftandMVeltmanAnnInstHenriPoincare20691974_AIHP,1974_Deser_OneLoopDivergencesofQuantizedEinsteinMaxwellFields_PRD}. This means a quantum theory of GR is nonpredictive. 
In the course of solving the renormalization problem, a class of fourth-order gravity theories \cite{1978StelleClassicalgravityhigherGRaG}, which improve on the renormalization behavior \cite{1977StelleRenormalizationhigherderivativePRD}, were investigated. Later is was shown that theories with six or more derivatives are even superrenormalizable \cite{1997AsoreyLopezShapiroSomeremarkshighIJoMPA,2012_Modesto_SuperRenormalizableQuantumGravity_PRD,2013_Modesto_SuperRenormalizableMultidimensionalGravityTheoryandApplications_AR}.
Unfortunately, the (super)renormalization property comes along with the problem of Ostrogradsky instabilities \cite{1850_Ostrogradsky_MemoiresSurLesEquationsDifferentiellesRelativesAuProblemeDesIsoperimetres_MASP,2015_Woodard_OstrogradskysTheoremonHamiltonianInstability_S} manifesting as ghost fields, which are states with negative kinetic energy. The appearance of ghost states is a general feature of theories with higher derivatives and could be used as an argument to invalidate these theories. Nevertheless, as the (super)renormalizability seemed so promising, people were strongly motivated to resolve the ghost problem, e.g., by modifying the quantization scheme \cite{2000MannheimDavidsonFourthordertheoriesaph,2007BenderMakingsensenonRoPiP,2008BenderMannheimExactlysolvablePPRD,2008BenderMannheimGivingghostJoPAMaT,2008BenderMannheimNoghosttheoremPRL,2013MannheimPTsymmetryasPTotRSoLAMPaES,2018_Mannheim_AntilinearityRatherThanapa,2016MannheimExtensionCPTtheoremPLB,2016_Salvio_QuantumMechanicsof4DerivativeTheories_TEPJC,2017_Accioly_LowEnergyEffectsinaHigherDerivativeGravityModelwithRealandComplexMassivePoles_PRD} or by nonlocal theories \cite{1997_Tomboulis_SuperrenormalizableGaugeandGravitationalTheories_aph,2006_Biswas_BouncingUniversesinStringInspiredGravity_JoCaAP,2012_Biswas_TowardsSingularityandGhostFreeTheoriesofGravity_PRL,2014_Modesto_SuperRenormalizableandFiniteGravitationalTheories_NPB}.
But still there is no unanimous opinion about the ghost issue \cite{2017_Raidal_OntheQuantisationofComplexHigherDerivativeTheoriesandAvoidingtheOstrogradskyGhost_NPB}. Hence, we do not treat the ghost problem in this work.

However, even if it turns out that the ghost problem cannot be solved, it is interesting to consider theories with higher derivatives as effective theories in the low-energy regime of more fundamental theories, like string theory, which do not suffer from ghost instabilities and renormalization problems \cite{1985_Zwiebach_CurvatureSquaredTermsandStringTheories_PLB,1986_Deser_StringInducedGravityandGhostFreedom_PLB,1989_Jones_FieldRedefinitionDependenceoftheLowEnergyStringEffectiveAction_ZfuPCPaF}.

Many tests have been performed to constrain theories of modified gravity.
Besides the direct detection of gravitational waves by the aLIGO/VIRGO interferometers \cite{abbott2016observation,2016_Abbott_GW151226_Observationofgravitationalwavesfroma22-solar-massbinaryblackholecoalescence_PRL,2017_Scientific_GW170104_Observationofa50-Solar-MassBinaryBlackHoleCoalescenceatRedshift0.2_PRL,2017_Abbott_GW170608Observationofa19SolarMassBinaryBlackHoleCoalescence_TAJL,2017_Abbott_GW170814_Athree-detectorobservationofgravitationalwavesfromabinaryblackholecoalescence_PRL,2017_Abbott_GW170817_PRL},
one can use the indirect detection of gravitational waves by measuring the decrease of the orbital period of stellar binary systems. This decrease of the orbital period agrees to high precision with the prediction of GR \cite{2004WeisbergTaylorRelativisticbinarypulsarapa}, and hence it is ideal to test theories of modified gravity.
Besides that, the first detection of gravitational waves (GWs) from a binary neutron star merger GW170817 with electromagnetic follow-up signals coming from GRB 170817A \cite{2017_Abbott_GW170817_PRL,2017_Abbott_Multi-messengerobservationsofabinaryneutronstarmerger_AJL,2017_Coulter_SwopeSupernovaSurvey2017aSSS17atheOpticalCounterparttoaGravitationalWaveSource_S} in a large range of the energy spectrum put strong constraints on the speed of the gravitational waves and hence on many theories of modified gravity \cite{2017_Abbott_GravitationalWavesandGammaRaysfromaBinary,
2017_Lombriser_ChallengestoSelfAccelerationinModifiedGravity,
2017_Baker_StrongConstraintsonCosmologicalGravityfromGW170817andGRB170817A_Prl,2017_Creminelli_DarkEnergyAfterGW170817andGRB170817A_PRL,2017_Ezquiaga_DarkEnergyAfterGW170817DeadEndsandtheRoadAhead_PRL,
2017_Sakstein_ImplicationsoftheNeutronStarMergerGW170817forCosmologicalScalarTensorTheories_Prl,
2018_Nersisyan_GravitationalWaveSpeedImplicationsforModelswithoutaMassScale_,
2018_Akrami_NeutronStarMergerGW170817StronglyConstrainsDoublyCoupledBigravity_apa}.

In this work, we want to analyze aspects of the gravitational wave solutions in generalized higher-derivative gravity \cite{1997AsoreyLopezShapiroSomeremarkshighIJoMPA} (see also \cite{2017_Accioly_LowEnergyEffectsinaHigherDerivativeGravityModelwithRealandComplexMassivePoles_PRD, 2017_Accioly_OntheGravitationalSeesawinHigherDerivativeGravity_TEPJCa,2017_Giacchini_OntheCancellationofNewtonianSingularitiesinHigherDerivativeGravity_PLBa,2018_Giacchini_WeakFieldLimitandRegularSolutionsinPolynomialHigherDerivativeGravities_apa} for studies of measurable effects in the low-energy regime).
We structure this work in the following way. In Sec. \ref{basics}, we introduce the theory of higher-derivative gravity and derive the linearized field equations for the gravitational field containing a massive spin-0, a massless spin-2, and a massive spin-2 field (the massless spin-2 obeys the same field equation as in GR). After that, in Sec. \ref{solutions} using the method of Green's function, we analyze the massive spin-0 and the massive spin-2 fields in the special case of quadratic gravity. To see how these modes affect the gravitational radiation, we study the solutions in the presence of a binary system in circular motion and in the Newtonian limit. For the massive spin-2 field we also analyze the number of degrees of freedom (d.o.f.) that are excited by a matter source. Finally, we summarize and conclude.

Throughout the paper, we use $c = \hbar = 1$. Latin indices run from $1$ to $D$ and greek indices run from $0$ to $D$, where $D$ is the number of spacetime  dimensions. Repeated indices are implicitly summed over. $\mathbf{x}$ denotes the $(D~-~1)$-dimensional spatial vector. Further conventions are defined in Appendix \ref{conventions}.

\section{Fourth-Order Derivative Gravity \label{basics}}

In this work we study a general class of higher-derivative theories of gravity, which are invariant under general coordinate transformations and include terms up to quadratic order in curvature tensors. In such a case the most general $D$-dimensional $(D \geq 3)$ action is given by \cite{1997AsoreyLopezShapiroSomeremarkshighIJoMPA}
\begin{align}
S = \!\!\int\!\! d^D x \frac{\sqrt{-g}}{64\pi G}&\left[-4\epsilon R + R F_1(\Box)R + R_{\mu\nu} F_2(\Box) R^{\mu\nu}\right.\nonumber \\
&\,\,+ \left. R_{\mu\nu\rho\sigma} F_3(\Box) R^{\mu\nu\rho\sigma} \right]\, +\, S_m,
	\label{action}
\end{align}
where $G$ is Newton's constant, $g = \text{det}(g_{\mu\nu})$ is the determinant of the metric and $S_m$ is the matter action. 
$R^\mu_{\nu\rho\sigma}$, $R_{\mu\nu}$ and $R$ are the Riemann tensor, Ricci tensor and Ricci scalar defined in Appendix \ref{conventions}. $\epsilon$ is a parameter that takes the values $\pm 1$. $F_1(\Box)$, $F_2(\Box)$ and $F_3(\Box)$ are polynomial functions of the covariant d'Alembert operator $\Box = g^{\mu\nu} \nabla_\mu \nabla_\nu$ defined by $F_i(\Box) = \sum_{k = 0}^{n} \alpha_{i_k} \Box^k$, where $n$ is an arbitrary but finite nonnegative integer and $\alpha_{i_k}$ are real coefficients with a canonical mass dimension $M^{-2(k + 1)}$. $\Box^k = \left( \nabla_\rho \nabla^\rho \right)^k$ acts on tensors in the usual sense. If $F_1$, $F_2$, and $F_3$ are independent of the d'Alembert operator (corresponding to $n = 0$), \eqref{action} reduces to the action of quadratic gravity (see, e.g., \cite{2003DeserEnergygenerichigher,2010NelsonStaticsolutionsfourth}). As a special case of quadratic gravity, for $D = 4$, $\epsilon = -1$, $F_1 = 128\pi G\alpha_g/3$, $F_2 = -128\pi G\alpha_g $, and $F_3 = 0$ the action reduces to conformal gravity \cite{2018_Caprini_AstrophysicalGravitationalWavesinConformalGravity_apa}. Note also that for $D = 4$, $\epsilon = +1$, and $F_1 = F_2 = F_3 = 0$ one recovers the Einstein-Hilbert action.

\subsection{Linearized Wave Equations}

We are interested in the linearized version of this theory in a flat Minkowski background spacetime and hence we write 
\begin{equation}
g_{\mu\nu} = \eta_{\mu\nu} + h_{\mu\nu},
\end{equation}
where $h_{\mu\nu}$ represents a small metric perturbation and the d'Alembert operator reduces to $\Box = \eta^{\mu\nu}\partial_\mu\partial_\nu$.

In linearized theory there is a useful relation \cite{J.deAlmeida_TeyssandierlikeGaugeConditionforaGeneralClassofHigherDerivativeGravity}
\begin{align}
R_{\mu\nu\rho\sigma}F_3(\Box) R^{\mu\nu\rho\sigma} \!=\, &4 R_{\mu\nu} F_3(\Box) R^{\mu\nu} - R F_3(\Box) R + \partial\Omega\nonumber\\ 
&+ \mathcal{O}(h^3),
	\label{relation between curvature terms}
\end{align}
where $\partial\Omega$ denotes a surface term, which does not contribute to the field equations for appropriate boundary conditions. We also neglect terms cubic in the metric perturbation, since these terms do not contribute to the linearized field equations. This shows that by a redefinition of $F_1(\Box)$ and $F_2(\Box)$ we can set $F_3(\Box) = 0$. For this reason we drop the $F_3$-term in the further analysis. 

Expanding \eqref{action} to second order in $h_{\mu\nu}$ (for $F_3 = 0$) and varying with respect $h_{\mu\nu}$, we find the linearized field equations
\begin{align}
&\left(\epsilon + \frac{1}{4} F_2(\Box)\Box\right)G_{\mu\nu}^{(1)} + \left(\frac{1}{2}F_1(\Box) + \frac{1}{4} F_2(\Box) \right)\nonumber\\
&\times\left( \eta_{\mu\nu}\Box - \partial_\mu\partial_\nu \right)R^{(1)} = - 8\pi G T_{\mu\nu},
	\label{eom metric}
\end{align}
where $G_{\mu\nu} = R_{\mu\nu} - 1/2\eta_{\mu\nu}R$ is the Einstein tensor and $\ldots^{(1)}$ denotes quantities that are linear in $h_{\mu\nu}$. We observe that the field equations do not contain derivatives of the $F_i$'s. This is a result of the linearization because covariant derivatives reduce to partial derivatives which commute with the $F_i$'s. This is not true for the nonlinear field equations (for details, see \cite{2013BiswasGeneralizedghostfree}). For a list of the linearized curvature tensors, see Appendix \ref{conventions}. The matter energy-momentum tensor is defined by
\begin{equation}
T_{\mu\nu} = -\frac{2}{\sqrt{-g}}\frac{\delta S_m}{\delta g^{\mu\nu}}.
\end{equation}
The trace of \eqref{eom metric} is given by
\begin{align}
&\left(\frac{1}{2}F_1(\Box) + \frac{1}{4} F_2(\Box) \right)\Box R^{(1)}\nonumber\\
&\;\;\;= -\frac{8\pi G}{(D-1)} T + \frac{D - 2}{2(D-1)}\left(\epsilon + \frac{1}{4}F_2(\Box)\Box \right) R^{(1)},
	\label{trace eom metric}
\end{align}
where $T = \eta^{\mu\nu} T_{\mu\nu}$ is the trace of the matter energy-momentum tensor.
Using \eqref{trace eom metric} one can rewrite \eqref{eom metric} as
\begin{align}
&\left( \epsilon + \frac{1}{4} F_2(\Box)\Box \right) \left( R_{\mu\nu}^{(1)} - \frac{1}{2(D-1)}\eta_{\mu\nu} R^{(1)}\right)\nonumber\\
&\;\;\;- \left(\frac{1}{2}F_1(\Box) + \frac{1}{4} F_2(\Box) \right) \partial_\mu \partial_\nu R^{(1)}\nonumber\\
&= -8\pi G\left(T_{\mu\nu} - \frac{1}{D-1} \eta_{\mu\nu}T \right).
	\label{eom metric 2}
\end{align}
Now, it is convenient to define 
\begin{align}
Z_\mu \equiv &-\left( \epsilon + \frac{1}{4} F_2(\Box)\Box \right) \partial^\alpha \bar{h}_{\alpha\mu}\nonumber\\  
&- \left( \frac{1}{2}F_1(\Box) + \frac{1}{4} F_2(\Box) \right) \partial_\mu R^{(1)},
	\label{Tgauge}
\end{align}
where $\bar{h}_{\mu\nu} = h_{\mu\nu} - 1/2 \eta_{\mu\nu} h$ is the trace-reversed metric perturbation and $h = \eta^{\mu\nu}h_{\mu\nu}$.
Using \eqref{Tgauge} we can bring \eqref{eom metric 2} to the form
\begin{align}
&\left( \epsilon + \frac{1}{4} F_2(\Box)\Box \right) \left( \frac{1}{2} \Box h_{\mu\nu} - \frac{1}{2(D-1)} \eta_{\mu\nu} R^{(1)} \right)\nonumber\\
&+ \!\frac{1}{2}\! \left(\partial_\nu Z_\mu \!+\! \partial_\mu Z_\nu \right) = -8\pi G \left( T_{\mu\nu} - \frac{1}{D - 1}\eta_{\mu\nu} T\right).
	\label{eom metric 3}
\end{align}

Making use of the invariance under infinitesimal coordinate transformations, $x^\mu \rightarrow x^{\mu} + \xi^\mu$, where $|\partial_\mu\xi^\nu|$ is of the same order as $|h_{\mu\nu}|$, we choose the generalized Teyssandier gauge condition $Z_\mu = 0$ \cite{2017_Accioly,J.deAlmeida_TeyssandierlikeGaugeConditionforaGeneralClassofHigherDerivativeGravity}.
Hence, \eqref{eom metric 3} becomes
\begin{align}
&	\left( \epsilon + \frac{1}{4} F_2(\Box)\Box \right) \left( \frac{1}{2} \Box h_{\mu\nu} - \frac{1}{2 (D - 1)} \eta_{\mu\nu} R^{(1)} \right)\nonumber\\ 
&= -8\pi G \left( T_{\mu\nu} - \frac{1}{D-1} \eta_{\mu\nu} T \right).
	\label{eom metric Tgauge}
\end{align}

Opposed to GR, the metric perturbation contains more d.o.f. than just a massless spin-2 field. Therefore, it turns out to be convenient to write the metric perturbation as
\begin{equation}
h_{\mu\nu} = \epsilon(\eta_{\mu\nu} \phi + H_{\mu\nu} + \Psi_{\mu\nu}),
	\label{metric splitted}
\end{equation}
where $\phi$ resembles a massive spin-0 field, $H_{\mu\nu}$ is a massless spin-2 field and $\Psi_{\mu\nu}$ is a massive spin-2 field. The distinction between $H_{\mu\nu}$ and $\Psi_{\mu\nu}$ will become clear in the following.

Inserting \eqref{metric splitted} into \eqref{eom metric Tgauge} and following the steps in \cite{2017_Accioly,J.deAlmeida_TeyssandierlikeGaugeConditionforaGeneralClassofHigherDerivativeGravity} one finds
\begin{subequations}
\begin{align}
\left[ \Box - \epsilon m_\phi^2(\Box)\right]\phi &=  \frac{16\pi G}{(D - 1)(D - 2)}T,
	\label{massive spin0}
\end{align}
where the scalar field is defined as
\begin{equation}
\phi \equiv \frac{1}{(D - 1)m_\phi^2(\Box)} R^{(1)},
\end{equation}
and
\begin{equation}
m_\phi^2(\Box) \equiv \frac{4(D-2)}{4(D-1)F_1(\Box) + D F_2(\Box)}.
\end{equation}
\end{subequations}

The field equations and the gauge condition for the massless spin-2 field are
\begin{subequations}
\begin{align}
\Box \bar{H}_{\mu\nu} &= -16\pi G T_{\mu\nu},
	\label{massless spin2}\\
\partial^\mu \bar{H}_{\mu\nu} &= 0,
	\label{massless spin2 gauge}
\end{align}
\end{subequations}
where $\bar{H}_{\mu\nu} = H_{\mu\nu} - 1/2 \eta_{\mu\nu} H$ and $H = \eta^{\mu\nu}H_{\mu\nu}$.
Note that \eqref{massless spin2} and \eqref{massless spin2 gauge} lead to
\begin{equation}
\partial^\mu T_{\mu\nu} = 0
	\label{matter energy-momentum conservation}
\end{equation}
to lowest order in $h_{\mu\nu}$.

For the massive spin-2 field one finds
\begin{subequations}
\begin{align}
\left[ \Box - \epsilon m_\Psi^2(\Box) \right]\Psi_{\mu\nu} &= 16\pi G \left( T_{\mu\nu} - \frac{1}{D - 1} \eta_{\mu\nu} T \right),
	\label{massive spin2}\\
\partial_\mu\partial_\nu \Psi^{\mu\nu} &= \Box \Psi,
	\label{massive spin2 gauge}
\end{align}
where the massive spin-2 field is defined as
\begin{equation}
\Psi_{\mu\nu} \equiv \frac{1}{ m_\Psi^2} \left( \Box h_{\mu\nu} - m_\phi^2 \eta_{\mu\nu} \phi \right)
	\label{massive mode}
\end{equation}
and
\begin{equation}
m_\Psi^2(\Box) \equiv -\frac{4}{F_2(\Box)}.
	\label{graviton mass}
\end{equation}
\end{subequations}
It is useful to rewrite \eqref{massive spin2} and \eqref{massive spin2 gauge} as 
\begin{subequations}
\begin{align}
\left[ \Box - \epsilon m_\Psi^2(\Box) \right]\hat{\Psi}_{\mu\nu} &= 16\pi G T_{\mu\nu},
	\label{massive spin2 hat}\\
\partial_\mu\partial_\nu \hat{\Psi}^{\mu\nu} &= 0,
	\label{massive spin2 gauge hat}
\end{align}
\end{subequations}
where $\hat{\Psi}_{\mu\nu} \equiv \Psi_{\mu\nu} - \eta_{\mu\nu}\Psi$ and $\Psi = \eta^{\mu\nu}\Psi_{\mu\nu}$.

The distinction between $H_{\mu\nu}$ and $\Psi_{\mu\nu}$ can be easily understood in the case of quadratic gravity, where $m_{\Psi}(\Box) = m_{\Psi}$ is independent of the d'Alembert operator and thus represents an honest mass term. In this case \eqref{massive spin2} reduces to a massive Klein-Gordon equation, which shows that $\Psi_{\mu\nu}$ represents a massive wave. In addition, \eqref{massless spin2 gauge} constrains $\bar{H}_{\mu\nu}$ by four conditions, whereas \eqref{massive spin2 gauge} leads to only one condition on $\Psi_{\mu\nu}$. Hence, as usual for a massive wave, $\Psi_{\mu\nu}$ carries three additional d.o.f.
The case we consider here represents a generalization of quadratic gravity because $m_\Psi(\Box)$ depends on the d'Alembert operator.

Besides that, we observe that for $m_\phi \rightarrow \infty$ and $m_\Psi \rightarrow \infty$ the spin-0 field and massive spin-2 field become nondynamical. Hence, only the massless spin-2 field represents a dynamical d.o.f. For $D = 4$ and $\epsilon = +1$ this represents the GR limit. Conformal gravity is reproduced by $D = 4$, $\epsilon = -1$, $m_\Psi(\Box) = m_\Psi$, and $m_\phi \to \infty$. In this case there is no propagating scalar field, but a massless and a massive spin-2 field.

\section{Solutions} \label{solutions}

Using the method of Green's function, the solutions to \eqref{massive spin0}, \eqref{massless spin2}, and \eqref{massive spin2 hat} can be written as
\begin{subequations}
\begin{align}
\phi &= \frac{16\pi G}{(D - 1)(D - 2)} \int \mathrm{d}^D\! x^\prime\, \mathcal{G}_{\phi}(x - x^\prime) T(x^\prime),
	\label{solution massive spin0}\\
\bar{H}_{\mu\nu} &= -16\pi G \int \mathrm{d}^D\! x^\prime\, \mathcal{G}_{H}(x - x^\prime) T_{\mu\nu}(x^\prime),\\
\hat{\Psi}_{\mu\nu} &= 16\pi G \int \mathrm{d}^D\! x^\prime\, \mathcal{G}_{\Psi}(x - x^\prime) T_{\mu\nu}(x^\prime).
	\label{massive mode solution}
\end{align}
\end{subequations}
The propagators $\mathcal{G}_{\phi}$, $\mathcal{G}_{H}$, and $\mathcal{G}_{\Psi}$ are defined by
\begin{subequations}
\begin{align}
\left[ \Box - \epsilon m_\phi^2(\Box) \right] \mathcal{G}_{\Phi}(x - x^\prime) &= \delta^{(D)}(x - x^\prime),\\
\Box \mathcal{G}_{H}(x - x^\prime) &= \delta^{(D)}(x - x^\prime),\\
\left[ \Box - \epsilon m_\Psi^2(\Box) \right] \mathcal{G}_{\Psi}(x - x^\prime) &= \delta^{(D)}(x - x^\prime).
	\label{definition of massive spin-2 propagator}
\end{align}
\end{subequations}

Contracting \eqref{massive mode solution} with a partial derivative yields
\begin{align}
\partial^\mu\hat{\Psi}_{\mu\nu}  = &\int_{V} d^D x^\prime \left(\frac{\partial}{\partial x_\mu} \mathcal{G}_\Psi(x-x^\prime)\right) T_{\mu\nu}(x^\prime)\nonumber\\
 = &- \int_V d^D x^\prime \left(\frac{\partial}{\partial x^\prime_\mu} \mathcal{G}_\Psi(x-x^\prime)\right) T_{\mu\nu}(x^\prime)\nonumber\\
 = &- \mathcal{G}_\Psi(x-x^\prime) T_{\mu\nu}(x^\prime)\vert_{\partial V}\nonumber\\
&+ \int_{V} d^D x^\prime  \mathcal{G}_\Psi(x-x^\prime) \left(\frac{\partial}{\partial x^\prime_\mu}T_{\mu\nu}(x^\prime)\right)\nonumber\\
= \,&0,
	\label{harmonic gauge for massive part}
\end{align}
where we have used $\frac{\partial}{\partial x^\mu} \mathcal{G}_\Psi(x-x^\prime)= - \frac{\partial}{\partial x^{\prime\mu}} \mathcal{G}_\Psi(x-x')$ for the second equal sign and integration by parts for the third equal sign. Furthermore, we have chosen a D-dimensional integration volume $V$ that is larger than the source, such that $T_{\mu\nu}(x)$ vanishes on its boundary $\partial V$. The last expression vanishes due to energy-momentum conservation given in \eqref{matter energy-momentum conservation}.

Using
\begin{equation}
\!\!\!\!\!\int\!\!\! \frac{\mathrm{d}^{D-1}k}{(2\pi)^{D-1}}g(|\mathbf{k}|)e^{i\mathbf{k} \cdot \mathbf{r}} 
\!\!=\!\!
\frac{1}{(2\pi)^{\!\frac{D-1}{2}} r^{\!\frac{D-3}{2}}}\!\!\! \int_0^\infty\!\!\!\!\!\!\! \mathrm{d}y y^{\frac{D-1}{2}} g(y) J_{\!\frac{D-3}{2}}(yr),
\end{equation}
where $g(x)$ is an arbitrary function and $J_n$ are the Bessel functions of the first kind. The frequency domain propagators for $D \geq 4$, given \eqref{solution massive spin0}-\eqref{massive mode solution}, can be written as
\begin{subequations}
\begin{align}
\tilde{\mathcal{G}}_\phi(\omega,\mathbf{x} - \mathbf{x^\prime}) 
		= &\frac{1}{(2\pi)^{\frac{D-1}{2}} \vert \mathbf{x} - \mathbf{x^\prime} \vert^{\frac{D-3}{2}}} \int_0^\infty \mathrm{d}k\nonumber\\ 
		&\times \frac{k^{\frac{D-1}{2}}}{{\omega^2 - k^2 - \epsilon m_\phi^2(\Box)}} J_{\frac{D-3}{2}} (k\vert  \mathbf{x} - \mathbf{x^\prime} \vert),
	\label{massive scalar green function}
\end{align}
\begin{align}
\tilde{\mathcal{G}}_H(\omega,\mathbf{x} - \mathbf{x^\prime})  
	= &\frac{1}{(2\pi)^{\frac{D-1}{2}} \vert \mathbf{x} - \mathbf{x^\prime} \vert^{\frac{D-3}{2}}} \int_0^\infty \mathrm{d}k\nonumber\\ 
	&\times \frac{k^{\frac{D-1}{2}}}{{\omega^2 - k^2}} J_{\frac{D-3}{2}} (k\vert  \mathbf{x} - \mathbf{x^\prime} \vert),
\end{align}
\begin{align}
\tilde{\mathcal{G}}_\Psi(\omega,\mathbf{x} - \mathbf{x^\prime}) 
	= &\frac{1}{(2\pi)^{\frac{D-1}{2}} \vert \mathbf{x} - \mathbf{x^\prime} \vert^{\frac{D-3}{2}}} \int_0^\infty \mathrm{d}k\nonumber\\ 
	&\times \frac{k^{\frac{D-1}{2}}}{{\omega^2 - k^2 - \epsilon m_\Psi^2(\Box)}} J_{\frac{D-3}{2}} (k\vert  \mathbf{x} - \mathbf{x^\prime} \vert).
	\label{massive tensor greens function}
\end{align}
\end{subequations}
Note that $m_\phi$ and $m_\Psi$ depend on the d'Alembert operator and hence on $\omega$ and $k$. Thus, we cannot calculate the $k$-integral in \eqref{massive scalar green function} and \eqref{massive tensor greens function} without specifying $F_1$ and $F_2$ or, equivalently, $m_\phi$ and $m_\Psi$.
To analyze the radiation behavior of these three fields, in the following we restrict to the case in which $F_1(\Box) = F_1$ and $F_2(\Box) = F_2$ are independent of the d'Alembert operator (and hence independent of $\omega$ and $k$). This represents the case of quadratic gravity. The massless spin-2 field is well known from GR and hence we only derive the solutions to the spin-0 and the massive spin-2 field equations.

\subsection{The Massive Spin-0 Field} \label{spin-0 field}

Inserting \eqref{massive scalar green function} into \eqref{solution massive spin0} for $D = 4$ yields
\begin{align}
\phi\left(t,\mathbf{x}\right) = &\frac{8\pi G}{3} \int \mathrm{d}^3\! x^\prime \int\frac{\mathrm{d}\omega}{2\pi} \int_{0}^{\infty} \frac{\mathrm{d}k}{(2\pi)^{3/2}}  \frac{k^{3/2}}{(\omega^2 - k^2 - \epsilon m_\phi^2)}\nonumber\\ 
&\times \frac{J_{1/2}(k\vert \mathbf{x} - \mathbf{x^\prime} \vert)}{\vert \mathbf{x} - \mathbf{x^\prime} \vert^{1/2}} \tilde{T}\left(\omega,\mathbf{x^\prime}\right),
	\label{spin0 field sol after angle integration}
\end{align}
where $m_\phi^2 = 2/(3F_1 + F_2)$ and $J_{1/2}(k\vert \mathbf{x} - \mathbf{x^\prime} \vert) = (2/\pi k\vert \mathbf{x}~-~\mathbf{x^\prime} \vert)^{1/2} \sin{(k\vert \mathbf{x} - \mathbf{x^\prime} \vert)}$.

For $\epsilon = +1$ and $m_\phi^2 > \omega^2$ (large mass of the massive spin-0 mode) the poles of \eqref{spin0 field sol after angle integration} are on the imaginary $k$-axis. This leads to an exponential suppression for the massive spin-0 mode and to a negligible contribution to the gravitational radiation (see \cite{2018_Caprini_AstrophysicalGravitationalWavesinConformalGravity_apa} for details). Hence, we do not study this case here. $\epsilon = -1$ and $m_\phi^2 > \omega^2$ leads to an oscillating gravitational potential, which we also do not want to study here.

For $m_\phi^2 < \omega^2$ (small mass of the massive spin-0 field) the poles are on the real $k$-axis. This leads to a massive propagating mode. Calculating the $k$-integral and using the quadrupole expansion ($k_{\omega,\phi} \mathbf{x^\prime} \cdot \mathbf{n} \ll 1$, where $k_{\omega,\phi} = \sqrt{\omega^2 - \epsilon m_\phi^2}$ and $\mathbf{n} = \mathbf{x}/r$, where $r = \vert \mathbf{x} \vert$, is the unit vector in $\mathbf{x}$-direction) in \eqref{spin0 field sol after angle integration}, the solution in the far-field (for $r \gg R$ we have $\vert \mathbf{x} - \mathbf{x^\prime} \vert \approx r - \mathbf{x^\prime} \cdot \mathbf{n} + O(R^2/r)$, where $R$ is the typical spatial scale of the source) is given by
\begin{widetext}
\begin{align}
\phi\left(t,\mathbf{x}\right) &= 
	-\frac{8\pi G}{3} \int \mathrm{d}^3\! x^\prime \int \frac{\mathrm{d}\omega}{2\pi} 			\frac{e^{ik_{\omega,\phi} \vert \mathbf{x} - \mathbf{x^\prime} \vert}\theta(\omega - m_\phi) + e^{-ik_{\omega,\phi} \vert \mathbf{x} - \mathbf{x^\prime} \vert}\,\theta(-\omega - m_\phi) }{4\pi\vert \mathbf{x} - \mathbf{x^\prime} \vert} \tilde{T}(\omega,\mathbf{x^\prime}) e^{-i\omega t} \nonumber \\	
	&= -\frac{G}{3 \pi r} \int \mathrm{d}^3\! x^\prime \left[ \int_{m_\phi}^{\infty} \mathrm{d}\omega e^{-i\omega t}   e^{ik_{\omega,\phi}r} \left( 1 - ik_{\omega,\phi} 			\mathbf{x^\prime} \cdot \mathbf{n} - \frac{k_{\omega,\phi}^2}{2}(\mathbf{x^\prime}\cdot			\mathbf{n})^2  \right) \tilde{T}(\omega,\mathbf{x^\prime}) + \int_{-\infty}^{-m_\phi} \!\!\!\!\ldots\;\; \right],
	\label{scalar field quadrupole approx}
\end{align}
\end{widetext}
where the $\int_{-\infty}^{-m_\phi}$-contribution will be suppressed in the steps below, because its analysis is the same as for the first integral.
The second line is exact up to the quadrupole and the far-field approximation.

It is convenient to define the mass-energy moments
\begin{subequations}
\begin{align}
M(t) &= 
\int \mathrm{d}^{3}\!x\, T^{00}(t,\mathbf{x}),\\
D^i(t) &= 
\int \mathrm{d}^{3}\!x\, x^{i} T^{00}(t,\mathbf{x}), \\
M^{ij}(t) &= 
\int \mathrm{d}^{3}\!x\, x^i x^j T^{00}(t,\mathbf{x}).
\end{align}
\end{subequations}
These quantities are called monopole, dipole, and quadrupole moments and in frequency space we denote them by $\tilde{M}(\omega)$, $\tilde{D}^i(\omega)$, and $\tilde{M}^{ij}(\omega)$.
Using these in \eqref{scalar field quadrupole approx} together with the relations \eqref{relation1}-\eqref{relation3} we get
\begin{align}
\phi\left(t,\mathbf{x}\right) \!=\! &-\frac{G}{3\pi r}\! \int_{m_\phi}^{\infty}\!\!\!\!\!	 \mathrm{d} \omega e^{-i\omega t} e^{i k_{\omega,\phi} r} \! \bigg(\!\!-\!\tilde{M}(\omega) \!+\! ik_{\omega,\phi} n_k \tilde{D}^k(\omega)\nonumber\\ 
&+ \frac{k_{\omega,\phi}^2}{2} n_k n_l \tilde{M}^{kl}(\omega) - \frac{\omega^2}{2} \tilde{M}_i^i(\omega) \bigg).
\end{align}
For simplicity, we make the assumptions that $m_{\phi}^2/\omega^2 \ll 1$.
Taking the time derivative and using $k_{\omega,\phi} \approx |\omega|(1 - \epsilon \frac{m_\phi^2}{2\omega^2})$ for $m_{\phi}^2/\omega^2 \ll 1$ leads to
\begin{align}
\dot{\phi}\left(t,\mathbf{x}\right) \!\approx\! &-\frac{G}{3\pi r} \int_{m_\phi}^{\infty}\!\!\!\!\! \mathrm{d} \omega\, e^{-i\omega t} e^{ i k_{\omega,\phi} r} \bigg(\!i \omega\tilde{M}(\omega) \!+\! \omega^2 n_k \tilde{D}^k(\omega)\nonumber\\ 
&- i \frac{\omega^3}{2} n_k n_l \tilde{M}^{kl}(\omega) + i \frac{\omega^3}{2} \tilde{M}_i^i(\omega) \bigg) ,
		\label{time derivative phi}
\end{align}
where the dot denotes the time derivative.

For a binary system with masses $m_1$ and $m_2$ on a circular orbit in the Newtonian limit, the contribution from the quadrupole moment can be described in the center of mass frame as originating from one particle with the reduced mass $\mu = m_1 m_2/(m_1 + m_2)$.
Assuming the orbit to be in the xy-plane, the nonvanishing components of the quadrupole moment  in the frequency domain are given by
\begin{subequations}
\begin{align}
\tilde{M}_{11}(\omega) \!&=\!
\frac{\mu R^{2}\pi}{2} \left[\delta\left(\omega\right) \!-\! \delta\left(\omega \!+\! 2\omega_{s}\right) \!-\! \delta\left(\omega \!-\! 2\omega_{s}\right)\right],
	\label{mass moment 11 omega}\\
\tilde{M}_{22}(\omega) \!&=\! \frac{\mu R^{2}\pi}{2} \left[\delta\left(\omega\right) \!+\! \delta\left(\omega \!+\! 2\omega_{s}\right) \!+\! \delta\left(\omega \!-\! 2\omega_{s}\right)\right],
	\label{mass moment 22 omega}\\
\tilde{M}_{12}(\omega) &= \frac{\mu R^{2}\pi}{2i} \left[\delta\left(\omega-2\omega_{s}\right) -\delta\left(\omega + 2\omega_{s}\right)\right],
	\label{mass moment 12 omega}\\
\tilde{M}^i_i(\omega) &= \mu R^{2}\pi \delta\left(\omega\right)
	\label{mass moment trace omega},
\end{align}
\end{subequations}
where $M_i^i = \delta^{ij} M_{ij}$ is the spatial trace of the quadrupole moment and $\omega_s > 0$ is the orbital frequency. 
The dipole moment can be written as
\begin{equation}
D^k = m x^k_{cm},
\end{equation}
where $m = m_1 + m_2$ is the total mass and $x^k_{cm} = (m_1 x^k_1 + m_2 x^k_2)/m$ is the $k$-th component of the center of mass coordinate. Hence, in the center of mass frame the dipole moment vanishes
\begin{equation}
\tilde{D}^k(\omega) = 0.
	\label{dipole moment}
\end{equation}
The monopole moment is given by
\begin{equation}
\tilde{M}(\omega) = m \delta(\omega).
	\label{monopole moment}
\end{equation}

Using \eqref{mass moment 11 omega}-\eqref{mass moment trace omega}, \eqref{dipole moment} and \eqref{monopole moment} in \eqref{time derivative phi}, we see that the monopole moment, the dipole moment and the trace of the quadrupole moment vanish and only the quadrupole contribution survives
\begin{equation}
\dot{\phi}\left(t,\mathbf{x}\right) \approx \frac{iG}{3\pi r} \int_{m_\phi}^{\infty} \mathrm{d} \omega\, e^{-i\omega t} e^{i k_{\omega,\phi} r} \frac{\omega^3}{2} n_k n_l \tilde{M}^{kl}(\omega).
	\label{spin-0 field quadr}
\end{equation}

The radiated energy in a three-dimensional volume $V$ larger than the source can be calculated by
\begin{equation}
\dot{E} = r^2\int_{\partial V} \!\!\mathrm{d}\Omega\,n_{s}T_{GRAV}^{s0},
	\label{radiated energy}
\end{equation}
where $\dot{E}$ is the time derivative of the gravitational energy, $T_{GRAV}^{\mu\nu}$ is the gravitational energy-momentum tensor and $d\Omega = \sin\theta\, d\theta\, d\phi$ is the differential solid angle.
In \eqref{spin-0 field quadr} we have shown that monopole and dipole moments do not contribute to time derivatives of the massive spin-0 field, but $T_{GRAV}^{s0}$ can also contain terms with spatial derivatives like $\partial^0\partial^s h_{\rho\sigma} \Box h^{\rho\sigma}$. By inserting \eqref{mass moment 11 omega}-\eqref{mass moment 12 omega} into \eqref{spin-0 field quadr} we can derive the relation
\begin{equation}
\partial^s \phi = \partial^0 \phi [1 - \epsilon \, m_\phi^2/(8\omega_s^2) + \mathcal{O}(m_\phi^4/\omega_s^4)] n^s + \mathcal{O}(1/r^2).
\end{equation}
This shows that spatial derivatives can be rewritten to time derivatives to leading order in $m_\phi^2/\omega_s^2$ and that monopole and dipole radiation do not contribute to \eqref{radiated energy}.

\subsection{The Massive Spin-2 Field}

In \eqref{harmonic gauge for massive part} we have shown that the harmonic gauge condition arises dynamically for the massive spin-2 field and hence in the presence of a source only the transverse modes of the massive spin-2 field are excited. Thus, we can use the residual gauge freedom to bring the massive spin-2 field to the transverse traceless (TT) gauge $\partial^\nu\Psi^{TT}_{\mu\nu} = 0, \Psi^{TT}_{0\mu} = 0 \text{ and, }\Psi^{TT} = 0$, meaning that the massless and massive spin-2 fields are constrained by the same number of conditions.  This fundamentally affects the gravitational radiation behavior in this model of higher derivative gravity. In vacuum quadratic gravity contains eight d.o.f.: one from the massive scalar field, two from the massless spin-2 field, and five from the massive spin-2 field. But GWs created by a source carry only five propagating d.o.f. since three of the five massive modes are not excited. Additionally, monopole and dipole radiation vanish.
To see this, we look at \eqref{massive mode solution} for $D = 4$ and $m_\Psi(\Box) = m_\Psi$, which is given by
\begin{align}
\hat{\Psi}_{\mu\nu}\left(t,\mathbf{x}\right) = &16\pi G\int \mathrm{d}^3\! x^\prime \int\frac{\mathrm{d}\omega}{2\pi} \int_{0}^{\infty} \frac{\mathrm{d}k}{(2\pi)^{3/2}}\nonumber\\
&\times \frac{k^{3/2}}{(\omega^2 - k^2 - \epsilon m_\Psi^2)} \frac{J_{1/2}(k\vert \mathbf{x} - \mathbf{x^\prime} \vert)}{\vert \mathbf{x} - \mathbf{x^\prime} \vert^{(1/2)}}\nonumber\\ 
&\times\tilde{T}_{\mu\nu}\left(\omega,\mathbf{x^\prime}\right),
	\label{multipole expansion h_solution_case1}
\end{align}
where $m_\Psi^2 = -4/F_2$ is independent of $\omega$ and $k$.

For the same reasons as for the spin-0 field we only study the case $m_\Psi^2 < \omega^2$ (small mass of the massive spin-2 field), which leads to a propagating wave and no oscillations in the gravitational potential.

Calculating the $k$-integral and using the quadrupole expansion ($k_{\omega,\Psi} \mathbf{x^\prime} \cdot \mathbf{n} \ll 1$, where $k_{\omega,\Psi} = \sqrt{\omega^2 - \epsilon m_\Psi^2}$) in \eqref{multipole expansion h_solution_case1}, the solution in the far field is given by
\begin{align}
\hat{\Psi}_{\mu\nu}\left(t,\mathbf{x}\right) = &-\frac{4 G}{r} \int \mathrm{d}^{3}\!x^\prime \bigg[ \int_{m_\Psi}^{\infty} \frac{\mathrm{d}\omega}{2\pi} e^{-i\omega t} e^{i k_{\omega,\Psi}r}\nonumber\\ 
&\times\left(1-ik_{\omega,\Psi}\mathbf{x^\prime} \cdot \mathbf{n}-\frac{k_{\omega,\Psi}^2}{2}\left(\mathbf{x^\prime} \cdot 
\mathbf{n}\right)^2\right)\nonumber\\ 
&\times\tilde{T}_{\mu\nu}\left(\omega,\mathbf{x^\prime}\right) + \int_{-\infty}^{-m_\Psi}\!\!\!\!\ldots\;\; \bigg] .
	\label{multipole expansion h_solution_case_expanded}
\end{align}
The second integral in the square brackets is suppressed in the steps below, because its analysis is analogous to the first integral. This expression is exact in the quadrupole approximation and the far-field approximation. 
Using \eqref{relation1}-\eqref{relation3} we can write the components of \eqref{multipole expansion h_solution_case_expanded} as
\begin{subequations}
\begin{align}
\hat{\Psi}^{00} = &-\frac{4 G}{r}\int_{m_\Psi}^{\infty} \frac{\mathrm{d}\omega}{2 \pi} e^{-i\omega t} e^{i k_{\omega,\Psi}r} \bigg(\tilde{M}(\omega)-ik_{\omega,\Psi}n_k \tilde{D}^{k}(\omega)\nonumber\\
&-\frac{k_{\omega,\Psi}^2}{2}n_k n_l \tilde{M}^{kl}(\omega)\bigg),
	\label{00_solution_fourier}\\
\hat{\Psi}^{0i} = &-\frac{4 G}{r}\int_{m_\Psi}^{\infty} \frac{\mathrm{d}\omega}{2 \pi} e^{-i\omega t} e^{ ik_{\omega,\Psi}r}\nonumber\\ 
&\times\left(-i\omega \tilde{D}^{i}(\omega)\times-\frac{\omega}{2}k_{\omega,\Psi}n_k \tilde{M}^{ki}(\omega)\right),
	\label{0i_solution_fourier}\\
\hat{\Psi}^{ij} = &\frac{2 G}{r}\int_{m_\Psi}^{\infty} \frac{\mathrm{d}\omega}{2 \pi}  e^{-i\omega t} e^{ ik_{\omega,\Psi}r} \omega^2 \tilde{M}^{ij}(\omega).
	\label{ij_solution_fourier}
\end{align}
\end{subequations}
We could use \eqref{dipole moment} to set the dipole contribution to zero, but it is instructive to keep it and to show that it also vanishes due to the dynamically induced gauge condition \eqref{harmonic gauge for massive part}.
Note that we can expand $k_{\omega,\Psi} \approx |\omega| \left(1-\epsilon\frac{m_\Psi^2}{2 \omega^2}\right)$ for $m_\Psi^2/\omega^2 \ll 1$. Using this expansion, the time derivatives of \eqref{00_solution_fourier}-\eqref{ij_solution_fourier} simplify to
\begin{subequations}
\begin{align}
\dot{\hat{\Psi}}^{00} \!\approx\! &-\frac{4 G}{r} \int_{m_\Psi}^{\infty}\!\! \frac{\mathrm{d}\omega}{2 \pi}\! e^{-i\omega t} e^{ik_{\omega,\Psi}r} \bigg(\!\! -i\omega \tilde{M}(\omega) - \omega^2 n_k \tilde{D}^k(\omega)\nonumber\\
&+ i \frac{\omega^3}{2} n_k n_l \tilde{M}^{kl}(\omega) \bigg),
	\label{time derivative massive wave 00}
\end{align}
\begin{align}
\dot{\hat{\Psi}}^{0i} \approx &-\frac{4 G}{r} \int_{m_\Psi}^{\infty} \frac{\mathrm{d}\omega}{2 \pi} e^{-i\omega t} e^{ik_{\omega,\Psi}r}\nonumber\\
&\times\left( -\omega^2 \tilde{D}^i(\omega) + i\frac{\omega^3 }{2} n_k \tilde{M}^{ki}(\omega) \right),
	\label{time derivative massive wave 0i}\\
\dot{\hat{\Psi}}^{ij} &\approx -i\frac{2 G}{r} \int_{m_\Psi}^{\infty} \frac{\mathrm{d}\omega}{2 \pi} e^{-i\omega t} e^{ik_{\omega,\Psi}r} \omega^3 \tilde{M}^{ij}(\omega).
	\label{time derivative massive wave ij}
\end{align}
\end{subequations}
Inserting \eqref{00_solution_fourier}-\eqref{ij_solution_fourier} explicitly into \eqref{harmonic gauge for massive part} leads to 
\begin{subequations}
\begin{align}
-i\int_{m_\Psi}^{\infty} \frac{\mathrm{d}\omega}{2\pi} e^{-i\omega t} e^{i k_{\omega,\Psi} r} \omega \tilde{M}(\omega) &= 0, 
	\label{hgauge1}\\
-\int_{m_\Psi}^{\infty} \frac{\mathrm{d}\omega}{2\pi} e^{-i\omega t} e^{i k_{\omega,\Psi} r}  \omega^2 \tilde{D}^i(\omega) &= 0.
	\label{hgauge2}
\end{align}
\end{subequations}
Using \eqref{hgauge1} and \eqref{hgauge2} in \eqref{time derivative massive wave 00}-\eqref{time derivative massive wave ij}, we see that monopole and dipole contributions vanish and only the quadrupole moment contributes. Note that this is a consequence of the conservation of the energy-momentum tensor in the linearized theory; see \eqref{matter energy-momentum conservation}. In contrast to GR, this result is approximate because in \eqref{time derivative massive wave 00}-\eqref{time derivative massive wave ij} we used $k_{\omega,\Psi} \approx \vert \omega \vert$.

Since the massive spin-2 field can be brought to the TT gauge and invoking \eqref{mass moment 11 omega}-\eqref{mass moment 12 omega}, the relevant components for the gravitational radiation are (note that we have to consider the $\int\limits_{-\infty}^{-m_\Psi}\!\!\!\!\ldots$ -contribution here)
\begin{align}
\hat{\Psi}_{11}(t,r) = -\hat{\Psi}_{22}\left(t,r\right) &= -\frac{4 G \mu R^{2}\omega_{s}^{2}}{r}\cos\left(2\omega_{s}t_{m}\right) ,
	\label{solution_MG_h11_c13} \\
\hat{\Psi}_{12}(t,r) = \hat{\Psi}_{21}(t,r) &= -\frac{4 G \mu R^{2}\omega_{s}^{2}}{r}\sin\left(2\omega_{s}t_{m}\right),
	\label{solution_MG_h12_c13}\\
\hat{\Psi}_i^i(t,r) &= \hat{\Psi}_{3i}(t,r) = \hat{\Psi}_{i3}(t,r) = 0,
	\label{solution_MG_trace_c13}
\end{align}
where $t_{m} = t - v_{m}r$ is the travel time and $v_{m} = \sqrt{1-\epsilon m_\Psi^2/(4\omega_s^2)}$ is the group velocity of the massive spin-2 field.
From this it becomes clear that spatial derivatives can be related to time derivatives by 
\begin{equation}
\partial^s \hat{\Psi}^{\rho \sigma} = \partial^0 \hat{\Psi}^{\rho \sigma} [1 - \epsilon \, m_\Psi^2/(8\omega_s^2) + \mathcal{O}(m_\Psi^4/\omega_s^4)] n^s + \mathcal{O}(1/r^2).
	\label{rel between time and spatial der}
\end{equation}
Hence, all spatial derivatives, which appear in the radiated energy, can be replaced by time derivatives to lowest order in $m_\Psi^2/\omega_s^2$. This demonstrates that no energy is carried away in monopole and dipole radiation by the massive spin-2 mode.

\section{Conclusion \label{conclusion}}

In this work we discussed the degrees of freedom and the gravitational radiation in generalized higher-derivative gravity and, in particular, in quadratic gravity. We derived the linearized field equations for the metric perturbation for $D$ dimensions and introduced the generalized Teyssandier gauge, which is convenient for higher-derivative theories. It turned out to be useful to separate the metric perturbation in a massive spin-0 field, a massless spin-2 field, and a massive spin-2 field, which obey massless and massive wave equations. We have shown that the massive spin-2 field satisfies the harmonic gauge condition, which originates from the conservation of the matter energy-momentum tensor in linearized theory. In Sec. \ref{solutions} we derived the solutions for the massive spin-0 and massive spin-2 field (the massless spin-2 field is well known from GR) by the methods of Green's function for $D = 4$ and constant masses, which represents the case of quadratic gravity. In this case the metric perturbation carries eight degrees of freedom in general. After that, to study the energy which is carried by these modes, we applied the solutions to a binary system in circular motion and in the Newtonian limit. For the massive spin-0 field and the massive spin-2 field it turned out that there is no monopole and dipole radiation, but only the quadrupole moment contributes. For the massive spin-2 field monopole and dipole radiation vanish as a consequence of the dynamically induced harmonic gauge condition. This means that, as for the massless spin-2 field, for the massive spin-2 field only the two transverse and traceless modes are excited by a matter source.

\begin{acknowledgments}

The author wishes to thank Dominik J. Schwarz for valuable discussions and suggestions on improving the manuscript, as well as Breno L. Giacchini and Gustatavo P. de Brito for useful hints on the literature. We acknowledge financial support from Deutsche
Forschungsgemeinschaft (DFG) under Grant No. RTG 1620 "Models of Gravity".
We also thank the COST Action CA15117 "Cosmology
and Astrophysics Network for Theoretical Advances and Training Actions (CANTATA)", 
supported by COST (European Cooperation in Science and Technology).
\end{acknowledgments}

%

\appendix

\section{Conventions \label{conventions}}

The signature of the metric is 
\begin{equation}
g = \text{diag}\left( -,+,+,+,\cdots,+ \right)
\end{equation}
containing one negative and $D-1$ positive entries.
The Christoffel symbols are defined by 
\begin{equation}
\Gamma_{\kappa\mu}^{\lambda} = 
\frac{1}{2}g^{\lambda\rho} \left(\partial_{\kappa}g_{\rho\mu} + \partial_{\mu}g_{\rho\kappa} - 
\partial_{\rho}g_{\kappa\mu}\right)
\end{equation}
and the Riemann tensor is given by 
\begin{equation}
R_{\mu\nu\kappa}^{\lambda} = 
-\left(\partial_{\nu}\Gamma_{\mu\kappa}^{\lambda} - 
	\partial_{\kappa}\Gamma_{\mu\nu}^{\lambda} + 
	\Gamma_{\nu\alpha}^{\lambda}\Gamma_{\mu\kappa}^{\alpha} - 
	\Gamma_{\kappa\alpha}^{\lambda}\Gamma_{\mu\nu}^{\alpha}\right).
\end{equation}
Contracting the first and the third index, we find the Ricci tensor
\begin{equation}
R_{\mu\nu} = R^\rho_{\mu\rho\nu}.
\end{equation}
The Ricci scalar is defined as
\begin{equation}
R = g^{\mu\nu} R_{\mu\nu}.
\end{equation}
The Einstein equations in the convention used in this work read
\begin{equation}
G_{\mu\nu} \equiv R_{\mu\nu} - \frac{1}{2}g_{\mu\nu}R =
 -8\pi G T_{\mu\nu} + \Lambda g_{\mu\nu}.
\end{equation}
A list of the curvature tensors to first
order in $h_{\mu\nu}$ is given by
\begin{align}
R_{\nu\rho\sigma}^{\mu\left(1\right)} & = \frac{1}{2}\left(-\partial_{\nu}\partial_{\rho}h_{\sigma}^{\mu}-\partial^{\mu}\partial_{\sigma}h_{\nu\rho}+\partial^{\mu}\partial_{\rho}h_{\nu\sigma}+\partial_{\nu}\partial_{\sigma}h_{\rho}^{\mu}\right),\\
R_{\mu\nu}^{\left(1\right)} & = \frac{1}{2}\left(\boxempty h_{\mu\nu}-\partial_{\rho}\partial_{\mu}h_{\nu}^{\rho}-\partial_{\nu}\partial_{\rho}h_{\mu}^{\rho}+\partial_{\mu}\partial_{\nu}h\right),\\
R^{\left(1\right)} & = \boxempty h-\partial_{\mu}\partial_{\nu}h^{\mu\nu}.
\end{align}

Further, we present useful relations between the energy-momentum tensor and the mass-energy moments using energy-momentum conservation in flat spacetime
\begin{align}
\int \mathrm{d}^D x \tilde{T}^{ij} \left(\omega,\mathbf{x}\right) &= -\frac{\omega^2}{2} \int \mathrm{d}^D x x^i x^j\tilde{T}^{00}(\omega,\mathbf{x})\nonumber\\ 
	&= -\frac{\omega^2}{2}\tilde{M}^{ij} \left(\omega\right),
	\label{relation1}\\
\int \mathrm{d}^D x\, \tilde{T}^{0i}(\omega,\mathbf{x}) &= -i\omega\int \mathrm{d}^D x\, x^i \tilde{T}^{00}(\omega,\mathbf{x}) \nonumber\\
	&= -i\omega \tilde{D}^i(\omega),\\
\int \mathrm{d}^D x\, \tilde{T}^{ij}(\omega,\mathbf{x}) &= -i\omega \int \mathrm{d}^D x\, x^i \tilde{T}^{j0}(\omega,\mathbf{x})\nonumber\\
	 &= -\frac{\omega^2}{2} \tilde{M}^{ij}(\omega).
	\label{relation3}
\end{align}

\end{document}